\documentclass[preprint,showpacs,amsmath,amssymb]{revtex4}
\usepackage{amssymb}

\usepackage{graphicx}
\usepackage{subfig}
\usepackage{tabularx}
\usepackage{dcolumn}
\usepackage{bm}

\begin{document}

\title{$\Sigma$ Resonances from $K^- N\rightarrow \pi\Lambda$
reactions with a center of mass energy from 1550 to 1676 MeV}

\author{ Puze Gao$^{a}$, Jun Shi$^{a}$, B.~S.~Zou$^{a,b}$}

\affiliation{a) Institute of High Energy Physics and Theoretical
Physics Center for Science Facilities, Chinese Academy of Sciences,
Beijing 100049, China\\
b) State Key Laboratory of Theoretical Physics, Institute of
Theoretical Physics, Chinese Academy of Sciences, Beijing 100190,
China}

\begin{abstract}

For the study of the $\Sigma$ resonances, we analyze the
differential cross sections and $\Lambda$ polarizations for the
reactions $K^-n\to\pi^-\Lambda$ and $K^-p\to\pi^0\Lambda$ with an
effective Lagrangian approach. Data of an early experiment and the
recent Crystal Ball experiment at BNL are included in the analysis
with the c.m. energy from 1550 to 1676 MeV. Our results clearly
support the existence of a $\Sigma$ resonance with $J^P={1\over
2}^+$, mass near 1633 MeV, and width about 120 MeV, which confirms
the 3-star $\Sigma(1660) \frac{1}{2}^+$ in PDG. Meanwhile, our
results do not support the existence of the 2-star
$\Sigma(1620){1\over 2}^-$ in PDG. The analysis results for the
parameters of the relevant $\Sigma$ resonances and couplings are
presented.
\end{abstract}
 \maketitle{}

\section{INTRODUCTION}

Quenched $qqq$ quark models and unquenched $qqq\leftrightarrow
qqqq\bar q$ quark models give very different predictions for the
$J^P={1\over 2}^-$ SU(3) nonet partners of the $N(1535)$ and
$\Lambda(1405)$. While quenched quark
models~\cite{Capstick0,capstick,gloz,loring} predict the
$J^P={1\over 2}^-$ $\Sigma$ and $\Xi$ resonances to be around 1650
MeV and 1760 MeV, respectively, the unquenched quark
models~\cite{pentq1,pentq2,zoupent} expect them to be around 1400
MeV and 1550 MeV, respectively, a meson-soliton bound-state approach
of the Skyrme model~\cite{Oh} and other meson-baryon dynamical
models~\cite{Kanchan11,Ramos} predict them to be around 1450 MeV and
1620 MeV, respectively.

Although various phenomenological models give distinguishable
predictions for the lowest $J^P={1\over 2}^-$ $\Sigma$ and $\Xi$
states, none of them are experimentally established. There is
relatively more information on the $\Sigma$ resonances in the PDG
tables, coming from analyses of early ${\overline K}N$ experiments
in the 1970s~\cite{pdg}. Some analyses are for the c.m. energy
around 1600 MeV~\cite{NPB94,NPB87,NPB109,PRD12}. However, restricted
by the uncertainties from low statistics and background
contributions, the $\Sigma$ resonant structures around 1600 MeV are
still not very clear, and several $\Sigma$ resonances are listed in
PDG tables with only one or two stars around this region.

There is a $\Sigma(1620){1\over 2}^-$ listed as a 2-star resonance
in the PDG tables~\cite{pdg}. This seems supporting the prediction
of quenched quark models. However, for the 2-star
$\Sigma(1620)\frac{1}{2}^-$ resonance, only four
references~\cite{16208,16207,16202,16201} are listed in PDG tables
with weak evidence for its existence. Among them, Ref.~\cite{16208}
and Ref.~\cite{16207} are based on multi-channel analysis of the
$\overline{K}N$ reactions.  Both claim evidence for a
$\Sigma(\frac{1}{2}^-)$ resonance with mass around 1620 MeV, but
give totally different branching ratios for this resonance.
Ref.~\cite{16208} claims that it couples only to $\pi\Lambda$ and
not to $\pi\Sigma$ while Ref.~\cite{16207} claims the opposite way.
Both analyses do not have $\Sigma(1660)\frac{1}{2}^+$ in their
solutions. However, Ref.~\cite{NPB94} shows no sign of
$\Sigma(\frac{1}{2}^-)$ resonance between 1600 and 1650 MeV through
analysis of the reaction $\overline{K}N\rightarrow\Lambda\pi$ with
the c.m. energy in the range of 1540-2150 MeV, instead it suggests
the existence of $\Sigma(1660)\frac{1}{2}^+$. Later multi-channel
analyses of the $\overline{K}N$ reactions support the existence of
the $\Sigma(1660)\frac{1}{2}^+$ instead of
$\Sigma(1620)\frac{1}{2}^-$~\cite{pdg}. In Ref.~\cite{16202}, the
total cross sections for $K^-p$ and $K^-n$ with all proper final
states are analyzed and indicate some $\Sigma$ resonances near 1600
MeV without clear quantum numbers. Ref.~\cite{16201} analyzes the
reaction $K^-n\rightarrow\pi^-\Lambda$ and gets two possible
solutions, with one solution indicating a $\Sigma({1\over 2}^-)$
near 1600 MeV, and the other showing no resonant structure below the
$\Sigma(1670)$. So all claims of evidence for the
$\Sigma(1620){1\over 2}^-$ listed in PDG tables~\cite{pdg} are very
shaky. Instead, some re-analyses of the $\pi\Lambda$ relevant data
suggest that there may exist a $\Sigma({1\over 2}^-)$ resonance
around 1380 MeV~\cite{Wujj}, which supports the prediction of
unquenched quark models~\cite{pentq1,pentq2}.

Some other works~\cite{NPB87,16206} show supports of $\Sigma({1\over
2}^-)$ with a larger mass, named as the $\Sigma(1750){1\over 2}^-$
in PDG. Ref.~\cite{NPB87} analyzes the reaction
$K^-p\rightarrow\pi^0\Lambda$ with the c.m. energy from 1537 to 2215
MeV, and gives possible $\Sigma(\frac{1}{2}^-)$ resonance with mass
around 1700 MeV. Ref.~\cite{16206} studies the same reaction with
the technique of Barrelet zeros for the partial wave solutions.
Seven ambiguous solutions are generated with several of them
containing $\Sigma({1\over 2}^-)$ with mass above 1650 MeV.

To pin down the nature of the lowest ${1\over 2}^-$ SU(3) baryon
nonet, it is crucial to find hyperon states of the lowest ${1\over
2}^-$ SU(3) nonet and study their properties systematically. For the
study of $\Sigma$ resonances, the ${\bar K} N\to\pi\Lambda$ reaction
is the best available channel, where the s-channel intermediate
states are purely hyperons with strangeness $S=-1$ and isospin
$I=1$.

Recently, high statistic new data for the reaction
$K^-p\rightarrow\pi^0\Lambda$ are presented by the Crystal Ball
collaboration with the c.m. energy of 1560-1676 MeV for both
differential cross sections and $\Lambda$ polarizations~\cite{CryB}.
Our previous analysis of the new Crystal Ball data clearly shows
that the Crystal Ball $\Lambda$ polarization data demand the
existence of a $\Sigma$ resonance with $J^P=\frac{1}{2}^+$ and mass
near 1635~MeV~\cite{pzgao}, compatible with
$\Sigma(1660)\frac{1}{2}^+$ listed in PDG, while the
$\Sigma(1620){1\over 2}^-$ is not needed by the data. The
differential cross sections alone cannot distinguish the two
solutions with either $\Sigma(1660)\frac{1}{2}^+$ or
$\Sigma(1620)\frac{1}{2}^-$.

In order to further clarify the status of the
$\Sigma(1620)\frac{1}{2}^-$ and the $\Sigma(1635)\frac{1}{2}^+$,
here we extend the work of Ref.~\cite{pzgao} to analyze the
differential cross sections and $\Lambda$ polarizations for both
$K^-p\to\pi^0\Lambda$ and $K^-n\to\pi^-\Lambda$ reactions with an
effective Lagrangian approach, using the new Crystal Ball data on
$K^-p\to\pi^0\Lambda$ with the c.m. energy of 1560-1676
MeV~\cite{CryB}, and the $K^-n\to\pi^-\Lambda$ data of
Ref.~\cite{16201} with the c.m. energy of 1550-1650~MeV, where the
evidence of the $\Sigma(1620)\frac{1}{2}^-$ was claimed.

This paper is organized as follows. In Section \ref{framework}, we
present the theoretical frame work of the analysis. In Section
\ref{results}, we present the analysis results and discussions. A
brief summary is given in section \ref{summary}.

\section{THEORETICAL FRAMEWORK} \label{framework}

The Feynman diagrams for ${K^-}N\rightarrow\pi\Lambda$ are depicted
in Fig.1, where $k$, $p$, $q$ and $p'$ represent the momenta of the
incoming $K^-$, $N$ and the outgoing $\pi$, $\Lambda$, separately.
The main contributions are from the t-channel $K^*$ exchange, the
u-channel proton exchange, and the s-channel $\Sigma$ and its
resonances.

\begin{figure}[htbp]
 \includegraphics*[width=8.5cm]{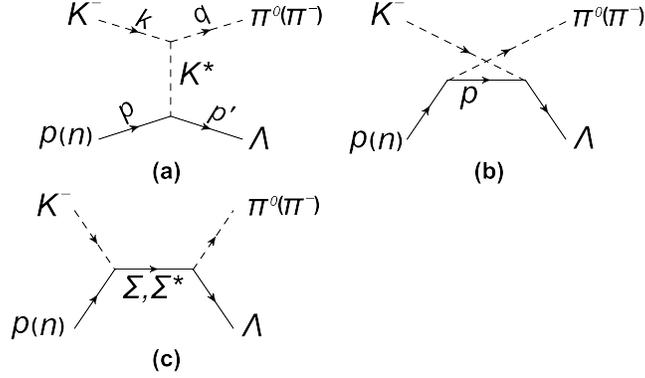}
\caption{Feynman diagrams for $K^-p\rightarrow\pi^0\Lambda$ and $K^-n\rightarrow\pi^-\Lambda$.
(a)t-channel $K^*$ exchange; (b)u-channel proton exchange;
(c)s-channel $\Sigma$ and its resonances exchanges.}\label{fig:graph}
\end{figure}

The relevant effective Lagrangians for the hadron couplings are
listed in Eq.(\ref{eq1}-\ref{eqn}). The value ranges of the coupling
constants or parameters are used exactly the same as those in
Ref.~\cite{pzgao}. Interested readers may refer to Ref.~\cite{pzgao}
for the detailed descriptions of our effective Lagrangians.

\begin{eqnarray}
  \mathcal{L}_{K^*K\pi}&=&i g_{K^*K\pi}K^*_\mu(\pi\cdot\tau\partial^\mu K-\partial^\mu \pi\cdot\tau K)\label{eq1}\\
 {\mathcal L}_{K^*N\Lambda}&=&-g_{K^*N\Lambda}\overline{\Lambda}(\gamma_\mu K^{*\mu} \nonumber\\
  &&-\frac{\kappa_{K^*N\Lambda}}{2M_N}\sigma_{\mu\nu}\partial^\nu K^{*\mu})N\\
  {\mathcal L}_{\pi NN}&=&\frac{g_{\pi NN}}{2M_N}\overline{N}\gamma^\mu\gamma_5\partial_\mu\pi\cdot\tau
  N\\
  {\mathcal L}_{KN\Lambda}&=&\frac{g_{KN\Lambda}}{M_N+M_\Lambda}\overline{N}\gamma^\mu \gamma_5 \Lambda\partial_\mu K +H.c.\\
  {\mathcal L}_{KN\Sigma(\frac{1}{2}^+)}&=&\frac{g_{KN\Sigma}}{M_N+M_\Sigma}\partial_\mu
  \overline{K}\overline{\Sigma}\cdot\tau\gamma^\mu\gamma_5 N+H.c.\\
  {\mathcal L}_{\Sigma(\frac{1}{2}^+)\Lambda\pi}&=&\frac{g_{\Sigma\Lambda\pi}}{M_\Lambda+M_\Sigma} \overline{\Lambda}\gamma^\mu\gamma_5\partial_\mu\pi\cdot\Sigma + H.c.\\
  {\mathcal L}_{KN\Sigma(\frac{1}{2}^-)}&=&-i g_{KN\Sigma}\overline{K}\overline{\Sigma}\cdot\tau N+H.c.\\
  {\mathcal L}_{\Lambda\pi\Sigma(\frac{1}{2}^-)}&=&-i g_{\Lambda\pi\Sigma}\overline{\Sigma} \Lambda\pi + H.c.\\
  {\mathcal L}_{KN\Sigma(\frac{3}{2}^+)}&=&\frac{f_{KN\Sigma}}{m_K}\partial_\mu\overline{K} \overline{\Sigma}^\mu\cdot\tau N + H.c.\\
  {\mathcal L}_{\Sigma(\frac{3}{2}^+)\Lambda\pi}&=&\frac{f_{\Sigma\Lambda\pi}}{m_\pi} \partial_\mu\overline{\pi}\cdot\overline{\Sigma}^\mu\Lambda + H.c.\\
  {\mathcal L}_{KN\Sigma(\frac{3}{2}^-)}&=&\frac{f_{KN\Sigma}}{m_K}\partial_\mu\overline{K} \overline{\Sigma}^\mu\cdot\tau\gamma_5 N + H.c.\\
  {\mathcal L}_{\Sigma(\frac{3}{2}^-)\Lambda\pi}&=&\frac{f_{\Sigma\Lambda\pi}}{m_\pi} \partial_\mu\pi\overline{\Sigma}^\mu\gamma_5\Lambda + H.c.\\
  {\mathcal L}_{KN\Sigma(\frac{5}{2}^-)}&=&g_{KN\Sigma}\partial_\mu\partial_\nu\overline{K} \overline{\Sigma}^{\mu\nu}\cdot\tau N+H.c.\\
  {\mathcal L}_{\Sigma(\frac{5}{2}^-)\Lambda\pi}&=&g_{\Lambda\pi\Sigma}\partial_\mu\partial_\nu \pi \cdot\overline{\Sigma}^{\mu\nu}\Lambda + H.c.\label{eqn}
\end{eqnarray}

Note that the isospin structures are contained in the Lagrangians,
e.g., the $K^*K\pi$ coupling is $\overline{K}^*\pi\cdot\tau K$ with
\begin{equation*}
  \overline{K}^*=(K^{*-},\overline{K}^{*0}), \pi\cdot\tau=\begin{pmatrix}
                                                            \pi^0 & \sqrt{2}\pi^+ \\
                                                            \sqrt{2}\pi^- & -\pi^0 \\
                                                          \end{pmatrix}
                                                          ,K=\begin{pmatrix}
                                                               K^+ \\
                                                               K^0 \\
                                                             \end{pmatrix};
\end{equation*}
and  for the $K N\Sigma$ coupling the isospin structure is
${\overline K}\overline{\Sigma}\cdot\tau N$ with
\begin{equation*}
   \overline{K}=(K^-,\overline{K}^0), \overline{\Sigma}\cdot\tau=\begin{pmatrix}
                                                            \overline{\Sigma}^0 & \sqrt{2}\overline{\Sigma}^+ \\
                                                            \sqrt{2}\overline{\Sigma}^- & -\overline{\Sigma}^0 \\
                                                          \end{pmatrix}
                                                          ,N=\begin{pmatrix}
                                                               p \\
                                                               n \\
                                                             \end{pmatrix}.
\end{equation*}

For each vertex of these channels, the following form factor is used
to describe the off-shell properties of the amplitudes:
\begin{equation}
  F_B(q^2_{ex},M_{ex})=\frac{\Lambda^4}{\Lambda^4 + (q^2_{ex}-M^2_{ex})^2},
\end{equation}
where $q_{ex}$ and $M_{ex}$ denote the 4-momenta and mass of the
exchanged hadron, respectively. The cutoff parameter $\Lambda$ is
constrained between 0.8 and 1.5~GeV for all channels.

For the propagators with 4-momenta $p$, we use
\begin{equation}
  \frac{-g_{\mu\nu}+p^\mu p^\nu/m^2_{K^*}}{p^2-m^2_{K^*}}
\end{equation}
for $K^*$ meson exchange ($\mu$ and $\nu$ are polarization index of $K^*$);
\begin{equation}
  \frac{\not\! p +m}{p^2-m^2}
\end{equation}
for spin-$\frac{1}{2}$ propagator;
\begin{equation}
  \frac{\not\! p +m}{p^2-m^2}(-g^{\mu\nu}+\frac{\gamma^\mu \gamma^\nu}{3}+\frac{\gamma^\mu p^\nu-\gamma^\nu p^\mu}{3 m}+\frac{2p^\mu p^\nu}{3 m^2})
\end{equation}
for spin-$\frac{3}{2}$ propagator; and
\begin{equation}
  \frac{\not\! p+m}{p^2-m^2}S_{\alpha\beta\mu\nu}(p,m)
\end{equation}
for spin-$\frac 5 2$ propagator, with
\begin{eqnarray}
  S_{\alpha\beta\mu\nu}(p,m) &=& \frac{1}{2}(\tilde{g}_{\alpha\mu}\tilde{g}_{\beta\nu}+ \tilde{g}_{\alpha\nu}\tilde{g}_{\beta\mu})-\frac{1}{5}\tilde{g}_{\alpha\beta} \tilde{g}_{\mu\nu} \nonumber\\
  & & -\frac{1}{10}(\tilde{\gamma}_\alpha\tilde{\gamma}_\mu \tilde{g}_{\beta\nu}+\tilde{\gamma}_\alpha\tilde{\gamma}_\nu \tilde{g}_{\beta\mu}\nonumber\\
  & &+\tilde{\gamma}_\beta\tilde{\gamma}_\mu  \tilde{g}_{\alpha\nu}+\tilde{\gamma}_\beta\tilde{\gamma}_\nu  \tilde{g}_{\alpha\mu}),
 \end{eqnarray}

 \begin{equation}
    \tilde{g}_{\mu\nu}=g_{\mu\nu}-\frac{p_\mu p_\nu}{m^2},~~~ \tilde{\gamma}_\mu
    =\gamma_\mu-\frac{p_\mu}{m^2}\not\! p.
 \end{equation}

For unstable resonances, we replace the denominator $1\over p^2-m^2$
in the propagators by the Breit-Wigner form $1\over
p^2-m^2+im\Gamma$, and replace $m$ in the rest of the propagators by
$\sqrt{p^2}$. The $m$ and $\Gamma$ in the propagators represent the
mass and total width of a resonance, respectively. Since all hyperon
states we include have rather narrow width with $\Gamma<<m$, the
pole positions for the states are basically $m - i \Gamma/2$.

The differential cross sections for $K^-N\rightarrow\pi\Lambda$ can
be expressed as
\begin{equation}
  \frac{d\sigma_{\pi\Lambda}}{d\Omega}=\frac{d\sigma_{\pi\Lambda}}{2\pi d \cos\theta}
  =\frac{1}{64 \pi^2 s}\frac{\left|\bf{q}\right|}{\left|\bf{k}\right|}\overline{\left|\mathcal M\right|}^2,
\end{equation}
where $\theta$ is the angle between the outgoing $\pi$ and the beam
direction in the c.m. frame; $s=(p+k)^2$, and $\bf{k}$ and $\bf{q}$
denote the 3-momenta of $K^-$ and $\pi$ in the c.m. frame,
respectively. And $\overline{\left|\mathcal M\right|}^2$ denotes the
spin averaged amplitude squared of the reaction.

The $\Lambda$ polarization in $K^- N\rightarrow\pi\Lambda\rightarrow\pi\pi N$ can be expressed as
\begin{equation}
 P_\Lambda=\frac{3}{\alpha_\Lambda}\left ( \int \cos\theta'\frac{d\sigma_{K^- N\rightarrow\pi\Lambda
 \rightarrow\pi\pi N}}{d\Omega d\Omega'}d\Omega'\right) \Big/\frac{d\sigma_{\pi\Lambda}}{d\Omega}
\end{equation}
where $\alpha_\Lambda=0.65$, and $d\Omega'=d \cos\theta'd\phi'$ is
the sphere space of the outgoing nucleon in the $\Lambda$ rest
frame, and $\theta'$ is the angle between the outgoing nucleon and
the vector $\bf{v}=\bf{k}\times\bf{q}$, which is perpendicular to
the $K^-N\rightarrow\pi\Lambda$ reaction plane.

For $\Lambda\rightarrow\pi N$, the effective Lagrangian is
\begin{equation}
  {\mathcal L}_{\Lambda\pi N}=G_F m_\pi^2\overline{N}(A-B\gamma_5)\Lambda,
\end{equation}
where $G_F$ is the Fermi coupling constant; $A$ and $B$ are
effective coupling constants.

The differential cross section for $K^- N\rightarrow\pi\Lambda\rightarrow\pi\pi N$ can be expressed as
\begin{equation}
  \frac{d\sigma_{K^- N\rightarrow\pi\Lambda\rightarrow\pi\pi N}}{d\Omega d\Omega'}= \frac{\left|\bf{q}\right|\left|\bf{p'}_N\right|\overline{\left|\mathcal M'\right|}^2}{32^2 2\pi m_\Lambda^2\Gamma_\Lambda s \left|\bf{k}\right|}
\end{equation}
where $\bf{p'}_N$ denotes the 3-momenta of the produced nucleon in
the $\Lambda$ rest frame, and $\Gamma_\Lambda$ is $\Lambda$ decay
width. ${\mathcal M}'$ denotes the amplitude of the reaction $K^-
N\rightarrow\pi\Lambda\rightarrow\pi\pi N$, and
$|\overline{\cal{M}'}|^2={1\over 2}\sum_{s_1,s_3}{\cal M'}{\cal
M'^+}$ is the spin averaged amplitude square.

\section{RESULTS and DISCUSSIONS}\label{results}
The isospin structures of the couplings require the cross section of
$K^-p\rightarrow\pi^0\Lambda$ to be half that of the
$K^-n\rightarrow\pi^-\Lambda$. In Fig.~2, We compare twice of the
Crystal Ball data~\cite{CryB} of the differential cross sections
with that of Ref.~\cite{16201} in similar beam momenta. And one can
see that in general, the data of the two experiments are compatible
with each other within statistic uncertainties.

\begin{figure}[htbp]
    {\includegraphics[width=8.6cm]{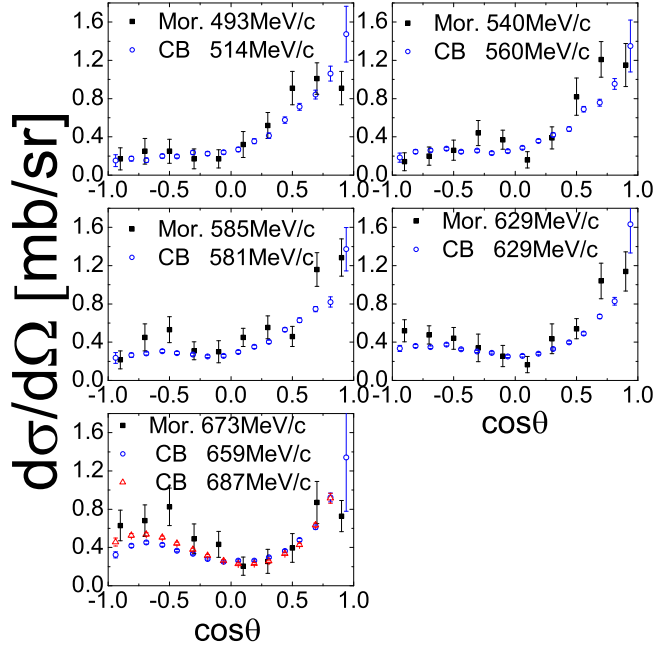}} 
  \caption{Comparison of the differential cross sections of Ref.~\cite{16201}
  with that of the Crystal Ball~\cite{CryB} of similar beam momenta, scaled with consideration of isospin relations. }
\end{figure}

In our analysis, the t-channel $K^*$ exchange and the u-channel
proton exchange amplitudes are fundamental ingredients. The well
established four-star $\Sigma(1189)\frac{1}{2}^+$,
$\Sigma(1385)\frac{3}{2}^+$, $\Sigma(1670)\frac{3}{2}^-$ and
$\Sigma(1775)\frac{5}{2}^-$ contributions are always included in the
analysis. The ranges of the parameters have been constrained from
the relevant PDG values or model predictions, which have been
explained in section II of Ref.~\cite{pzgao}. The mass of
$\Sigma(1775)\frac{5}{2}^-$ is much larger than the energy range of
the experiments, and the analysis is expected to be insensitive to
the parameters of $\Sigma(1775)$ resonance. Thus we fix the mass,
width and coupling constant of the $\Sigma(1775)$ to be the PDG
central values. The fixed parameters of $\Sigma(1189)\frac{1}{2}^+$,
$\Sigma(1385)\frac{3}{2}^+$ and $\Sigma(1775)\frac{5}{2}^-$ are
shown in Table \ref{table0} (other tunable parameters will be shown
in Table \ref{table1} and \ref{table2}).

\begin{table*}[htbp]
  \caption{fixed parameters for
$\Sigma(1189)\frac{1}{2}^+$, $\Sigma(1385)\frac{3}{2}^+$, and
$\Sigma(1775)\frac{5}{2}^-$.}
  \label{table0}
  \centering
    \begin{tabular}{c|c|c|c}
     \hline\hline
     \ & mass(MeV) & $\Gamma$(MeV) & $\sqrt{\Gamma_{\pi\Lambda}\Gamma_{\overline{K} N}}/\Gamma_{tot}$\\
     \hline
    $\Sigma(1189)\frac{1}{2}^+$ & $1192.6$  & $0$ & 
     \\
    \hline
    $\Sigma(1385)\frac{3}{2}^+$ & $1384$ & $36$ & 
    \\
    \hline
    $\Sigma(1775)\frac{5}{2}^-$ & $1775$ & $120$ &  $0.28$\\
     \hline\hline
    \end{tabular}
  \end{table*}

From analysis of the differential cross sections as well as the
$\Lambda$ polarizations of the two experiments with the above 6
channels and 14 tunable parameters constrained in the appropriate
ranges, we obtain a fit with $\chi^2$ of 1680 for the total 348 data
points. Here we only include the statistical errors presented by the
CB experiment. For the CB differential cross section data, there is
an overall systematical uncertainty of about 7\%.  Since the
systematical uncertainty for the CB data is mainly for the
normalization, the $\Lambda$ polarization defined by Eq.(23) does
not suffer such systematical uncertainty. Later we shall show that
for disentangling the ambiguity of spin-parity of $1/2^+$ or $1/2^-$
for an additional $\Sigma$ resonance it is mainly determined by
$\Lambda$ polarization data and hence does not suffer from such
systematical uncertainty.

To get an acceptable good fit of the experimental data, we need to
introduce some other $\Sigma$ resonances in s-channel, with its
coupling constant, mass and width as free parameters. Among the
$J^P=\frac{1}{2}^\pm,~~\frac{3}{2}^\pm$ $\Sigma$ resonances, we find
the best fit is given by including a $J^P=\frac{1}{2}^+$ resonance
with mass near 1633 MeV, width around 120 MeV, and couplings
$\sqrt{\Gamma_{\pi\Lambda}\Gamma_{\overline{K} N}}/\Gamma_{tot}\sim
-0.064$ where the negative sign means that the couplings to
$\pi\Lambda$ and $\bar KN$ have opposite signs.
 The analysis includes 18 tunable parameters in the allowed range
and the $\chi^2$ for this best fit is 572 for the total 348 data points. The
improvement is huge with $\Delta\chi^2=1008$ for 348 data points.

Fig.3 and Fig.4 show our analysis results for the differential cross
sections and the $\Lambda$ polarizations of the reaction compared
with the experimental data from Ref.~\cite{16201} and
Ref.~\cite{CryB}, respectively. We can see that the results are
generally in good agreement with the data, and the fit (especially,
the $\Lambda$ polarization in Fig.4) is much improved by including
the $\Sigma(1633){1\over 2}^+$. In Fig.3 we also show the second
solution of Ref.~\cite{16201}, which suggests a $\Sigma({1\over
2}^-)$ resonance with mass at 1600 MeV and width around 87 MeV. The
solution of Ref.~\cite{16201} fits the old data well, however, one
can see that the large error bars of the old data can accommodate
very different solutions, and the high precision new data of Crystal
Ball can distinguish different solutions more efficiently.

If we introduce the new resonance of other $J^P$ quantum numbers
instead of introducing the $\Sigma(1633){1\over 2}^+$ resonance, the
$\chi^2$ value is worse by 327 for $J^P=\frac{1}{2}^-$, 371 for
$J^P=\frac{3}{2}^+$, and 820 for $J^P=\frac{3}{2}^-$, respectively.
Our analysis with data of the two groups clearly supports the
existence of $\Sigma({\frac{1}{2}^+})$ resonance near 1633 MeV.
Further analysis with data from more groups and wider energy ranges
in the future will be helpful to verify our results.

\begin{figure}[htbp]
{\includegraphics*[width=15cm]{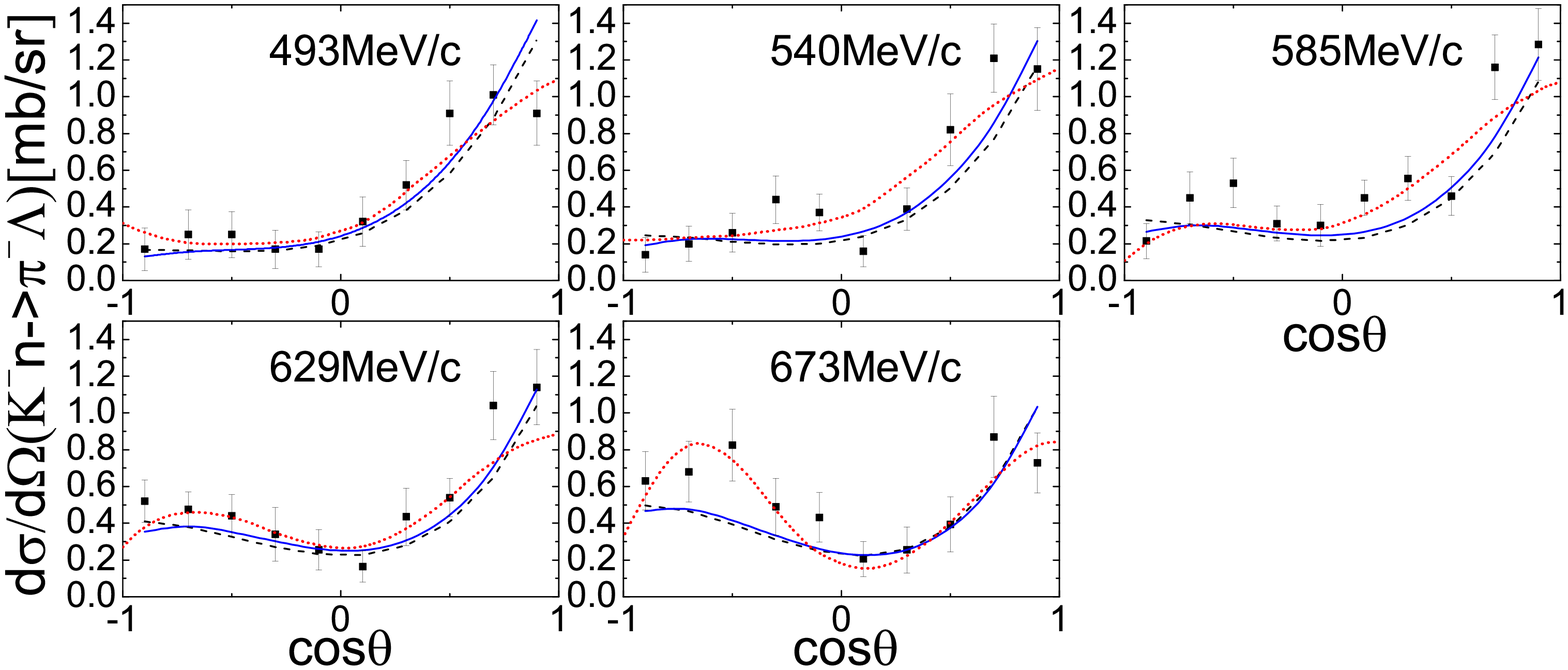}}
{\includegraphics*[width=15cm]{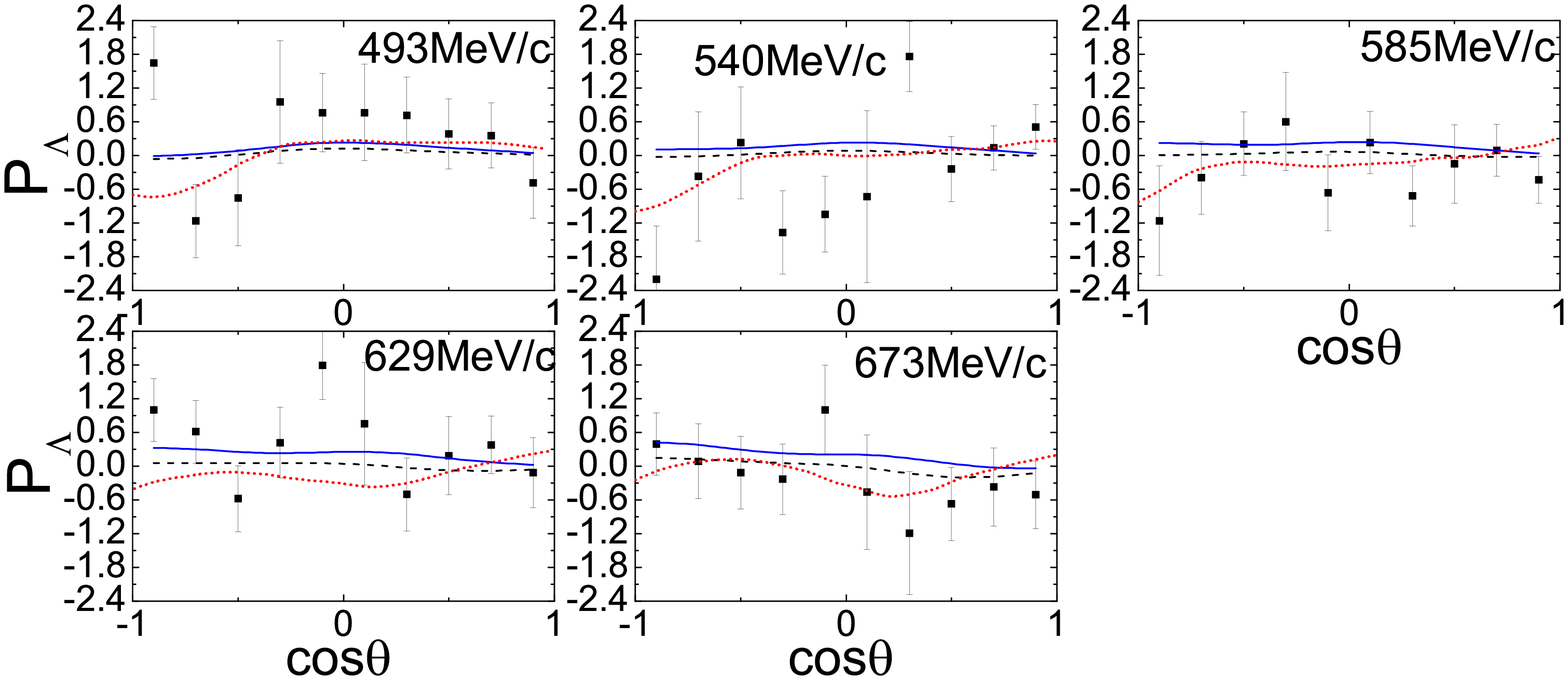}} \caption{(Color online)
The differential cross sections and $\Lambda$ polarizations of the
reaction $K^-n\rightarrow\pi^-\Lambda$, compared with the
experimental data of Ref.~\cite{16201} and its original second
solution including a $\Sigma({1\over 2}^-)$ resonance with mass
about 1600 MeV (red dotted lines), with incident $K^-$ momenta from
493 to 673 MeV/c in laboratory frame and $\theta$ the angle between
outgoing $\pi^-$ and incoming $K^-$ in the c.m. frame. The dashed
and solid (blue) lines are the best fits by including only the 4
well established $\Sigma$ resonances in s-channel, and by including
an additional $\Sigma(\frac{1}{2}^+)$ with mass around 1633 MeV,
respectively. }
\end{figure}

\begin{figure}[htbp]
{\includegraphics*[width=17cm]{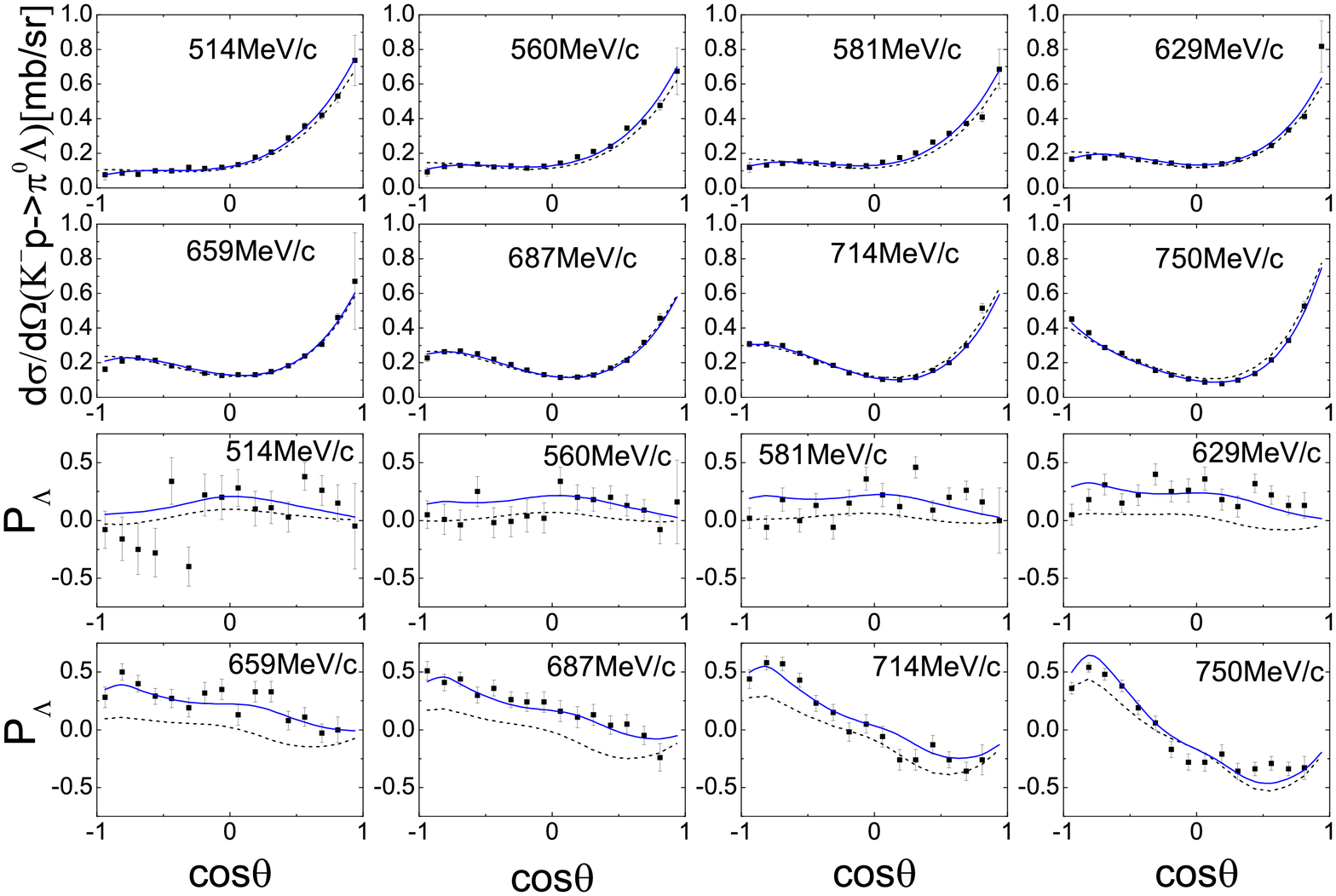}} \caption{(Color online) The
differential cross sections and $\Lambda$ polarizations for the
reaction $K^-p\rightarrow\pi^0\Lambda$, compared with the new
Crystal Ball data~\cite{CryB}, where $\theta$ is the angle between
outgoing $\pi^0$  and incoming $K^-$ in the c.m. frame. The dashed
and solid (blue) lines are the best fits by including only 4
established $\Sigma$ resonances in s-channel,  and by including an
additional $\Sigma(1633)\frac{1}{2}^+$ in s-channel, respectively.}
\end{figure}

In Table~\ref{table1}, we give the central values and uncertainties
for the 6 parameters of $\Sigma(1670)\frac{3}{2}^-$ and
$\Sigma(1633)\frac{1}{2}^+$ resonances. We can see that the mass and
width of the $\Sigma(1670)$ in our fit are compatible with the PDG
estimates~\cite{pdg}. The characters of $\Sigma({1\over 2}^+)$ from
our analysis are consistent with the 3-star
$\Sigma(1660)\frac{1}{2}^+$ in PDG, with more precise values for the
mass, width, and couplings.

\begin{table*}[htbp]
  \caption{Adjusted parameters for $\Sigma(1670)\frac{3}{2}^-$ and $\Sigma(1660)\frac{1}{2}^+$ resonances.}
  \label{table1}
  \centering
  \begin{tabular}{c|c|c|c}
     \hline\hline
    \ & mass(MeV)(PDG estimate) &
    $\Gamma_{tot}$(MeV)(PDG estimate) & $\sqrt{\Gamma_{\pi\Lambda}\Gamma_{\overline{K} N}}/\Gamma_{tot}$ (PDG range)\\
    \hline
    $\Sigma(1670)\frac{3}{2}^-$ & $1673\pm 1$(1665, 1685) & $52^{+5}_{-2}$(40, 80) & $0.081^{+0.002}_{-0.004}$(0.018,~0.17) \\
    \hline
    $\Sigma(1660)\frac{1}{2}^+$ & $1633\pm 3$(1630, 1690) & $121^{+4}_{-7}$(40, 200) & $-0.064^{+0.005}_{-0.003}$(-0.065, 0.24) \\
    \hline\hline
    \end{tabular}
\end{table*}

The other 12 tunable parameters in our study include 5 coupling
constants and 7 cut-off parameters. In Table~\ref{table2}, we show
the fitted results of the 5 coupling constants of the t-channel,
u-channel and s-channel $\Sigma(1189)$ and
$\Sigma(1385)\frac{3}{2}^+$ exchanges.

\begin{table*}[htbp]
  \caption{Adjusted coupling constants for t-channel, u-channel, and s-channel $\Sigma(1189)$ and $\Sigma(1385)\frac{3}{2}^+$ exchanges.}
  \label{table2}
  \centering
  \begin{tabular}{>{\scriptsize}c|>{\scriptsize}c|>{\scriptsize}c|>{\scriptsize}c|>{\scriptsize}c}
    \hline\hline
    $g_{K^*N\Lambda}$(model range) & $g_{K^*N\Lambda}\kappa_{K^*N\Lambda}$(model range) &
    $g_{\pi N N}g_{K N \Lambda}$(SU(3)) & $g_{K N \Sigma}g_{\Sigma\Lambda\pi}$(SU(3)) &
    $f_{K N \Sigma^*}f_{\Sigma^*\Lambda\pi}(SU(3))$ \\
    \hline
    $-6.10^{+0.07}_{-0}$(-6.11, -4.26)\cite{const} & $-11.33^{+0}_{-0.06}$(-16.3, -10.4)\cite{const} & $-178^{+2}_{-7}$(-176) & $49.2^{+0}_{-0.9}$(34.8) & $-3.94^{+0.32}_{-0.13}$(-4.1) \\
  \hline\hline
  \end{tabular}
 \end{table*}

All the fitted parameters listed in Table~\ref{table1} and
Table~\ref{table2} are consistent with those given in
Ref.~\cite{pzgao} within error bars.  The error bars listed here are
smaller than those of Ref.~\cite{pzgao}. The main reason is that we
made a careless mistake in Ref.~\cite{pzgao}: the output of
$\chi^2/2$ value was mistaken as $\chi^2$ value. The values of all
the $\chi^2$ in Ref.~\cite{pzgao} should be doubled. Another reason
is that here we include data of $K^-n\to\pi^-\Lambda$ reaction in
addition.

For the fit with $\Sigma(\frac{1}{2}^-)$  instead of including the
$\Sigma(1633){1\over 2}^+$, the fitted mass goes down to our preset
lower limit 1360 MeV, with width and coupling constant
$g_{KN\Sigma}g_{\Sigma\pi\Lambda}$  to be 312 MeV and -1.253,
respectively. So even in the case without including the
$\Sigma(1633){1\over 2}^+$,  the data prefer a low mass
$\Sigma(\frac{1}{2}^-)$ as indicated in Refs.~\cite{zoupent,Wujj},
rather than the $\Sigma(1620)\frac{1}{2}^-$ as listed in
PDG~\cite{pdg}.

When including both $\Sigma(1633)\frac{1}{2}^+$ and an additional
$\Sigma(\frac{1}{2}^-)$ in s-channel, we get a lowest $\chi^2$ of
548 for the total 348 data points with 22 tunable parameters. The
fitted mass and width of $\Sigma(\frac{1}{2}^-)$ are 1432 MeV and
$\geq 1000$ MeV, respectively.

From the above results, the $\Sigma(1620)\frac{1}{2}^-$ is not
supported from our analysis at all. This seems differing from the
results of Ref.~\cite{16201}, where one of its solutions supports
the $\Sigma(1620)\frac{1}{2}^-$, although another one of its
solutions does not need it. The major difference of two analyses is
the treatment of non-resonant background contribution. In
Ref.~\cite{16201}, ``a particular partial wave was assumed to be
either resonant or background but not both" and background
contributions in each partial waves are independent, while in our
approach the background contributions in each partial waves are
determined by the t-channel $K^*$ exchange and $u$-channel proton
exchange. If we only fit the data of Ref.~\cite{16201} with the
effective Lagrangian approach, with just the 4 established $\Sigma$
resonances in s-channel, together with the t-channel and u-channel
contributions, we obtain a $\chi^2$ of 116 for the total 100 data
points, which is already much smaller than the $\chi^2$ value of
$176\sim 180$ for both solutions of Ref.~\cite{16201}. We think our
approach is more physical and appropriate in describing the
reactions. Including the $\Sigma(1633)\frac{1}{2}^+$ in addition
will further reduce the $\chi^2$ to 109.  The old
$K^-n\to\pi^-\Lambda$ data with large error bars show marginal
evidence for the $\Sigma(1633)\frac{1}{2}^+$. It is mainly the new
precise Crystal Ball data on $\Lambda$ polarizations demanding the
existence of the $\Sigma(1633)\frac{1}{2}^+$.

Taking the four 4-star $\Sigma$ resonances and the 3-star
$\Sigma(\frac{1}{2}^+)$ with tunable parameters as basic
contributions, we further examine whether any other additional
resonance can make significant improvement to the fit.

The largest improvement is given by including an additional
$\Sigma(\frac{3}{2}^+)$ resonance with mass of 1840 MeV or above.
The $\chi^2$ reaches 487 for the total 348 data points, with the
resonance having a coupling $\sqrt{\Gamma_{\pi\Lambda}\Gamma_{\overline{K}
N}}/\Gamma_{tot}\sim 0.289$ and width around 271 MeV. Note there are
two $\Sigma(\frac{3}{2}^+)$ resonances above 1800 MeV listed in
PDG~\cite{pdg}, {\sl i.e.}, 1-star $\Sigma(1840)$ and  2-star
$\Sigma(2080)$. With this solution, the mass and width of the
$\Sigma({1\over 2}^+)$ shift to 1632 MeV and 93 MeV, respectively.

When including a $\Sigma(\frac{3}{2}^-)$ resonance, the best
$\chi^2$ is 535 for the total 348 data points, with resulted mass
around 1542 MeV, width about 25.6 MeV and couplings
$\sqrt{\Gamma_{\pi\Lambda}\Gamma_{\overline{K} N}}/\Gamma_{tot}\sim
0.0374$. This solution improves the $\chi^2$ by 37, and seems
consistent with the resonance structure $\Sigma(1560)$ or
$\Sigma(1580){3\over 2}^-$ in PDG. Ref.~\cite{D13} also proposes a
$\Sigma({3\over 2}^-)$ resonance with mass around 1570 MeV and width
about 60 MeV from $\overline{K}N\pi$ system. The inclusion of
$\Sigma(\frac{3}{2}^-)$ in the analysis makes the mass and width of
the $\Sigma(\frac{1}{2}^+)$ shift to 1634 MeV and 130 MeV,
respectively.

When including an additional $\Sigma(\frac{1}{2}^+)$ resonance, we
get a $\chi^2$ of 541 with mass 1610 MeV, width 20 MeV, and
$\sqrt{\Gamma_{\pi\Lambda}\Gamma_{\overline{K} N}}/\Gamma_{tot}\sim
-0.032$ for the additional $\Sigma(\frac{1}{2}^+)$. This resonance
seems consistent with the $\Sigma(1620)$ resonance with unknown
quantum numbers listed in PDG~\cite{pdg}. With the additional
$\Sigma(\frac{1}{2}^+)$ in analysis, the mass, width and
$\sqrt{\Gamma_{\pi\Lambda}\Gamma_{\overline{K} N}}/\Gamma_{tot}$ of
the $\Sigma(1633)\frac{1}{2}^+$ shift to 1647 MeV, 91 MeV and
-0.065. This solution appears to support the results of
Ref.~\cite{Kanchan08}, where both $\Sigma(1620){1\over2}^+$ and
$\Sigma(1660){1\over 2}^+$ are predicted. Although the $\chi^2$ is
only improved by 31 with the inclusion of two
$\Sigma(\frac{1}{2}^+)$ resonances at 1610 and 1647~MeV, compared
with the case of the single $\Sigma(1633){1\over{2}}^+$, the
existence of such two resonances can not be excluded. The
contribution of the $\Sigma(1633){1\over{2}}^+$ seems to have
similar effects with the two $\Sigma({1\over 2}^+)$ around with
narrower widths.

Some uncertainty may still exist from the uncertainties in some
coupling constants and cutoffs, however, the main results of this
analysis will not change.

\section{SUMMARY}\label{summary}

In order to further clarify the properties of the $\Sigma$
resonances, we analyze the differential cross sections and $\Lambda$
polarizations for the reactions $K^-n\rightarrow\pi^-\Lambda$ and
$K^-p\rightarrow\pi^0\Lambda$ with the effective Lagrangian method.
The experimental data are adopted from the new high statistic
Crystal Ball experiment~\cite{CryB} and an early report of
Ref.~\cite{16201}, with the c.m. energy in 1550-1676 MeV.

In our calculation, the contributions of the t-channel $K^*$
exchange, u-channel proton exchange and the four-star $\Sigma$
resonances exchanges in s-channel, $\it i.e.$, $\Sigma(1189)$,
$\Sigma(1385)$, $\Sigma(1670)$ and $\Sigma(1775)$ are always
included. These ingredients are still insufficient to describe the
experimental data, with $\chi^2$ about 1680 for the total 348 data
points.  An additional $\Sigma(\frac{1}{2}^+)$ with mass around 1633
MeV and width about 120 MeV is absolutely necessary to reach an
acceptable good fit. It reduces the $\chi^2$ to 572 for the total
348 data points. Its properties are consistent with the
$\Sigma(1660)\frac{1}{2}^+$ listed in PDG.

In searching for the lightest $\Sigma(\frac{1}{2}^-)$, our results
do not show any evidence for the $\Sigma(1620)\frac{1}{2}^-$
resonance listed as a 2-star resonance in PDG;  a
$\Sigma(\frac{1}{2}^-)$ with much lower mass as suggested by the
penta-quark models~\cite{pentq1,pentq2} cannot be excluded. The
indications of the other additional $\Sigma$ resonance structures
are also discussed, with possible $\Sigma({3\over 2}^-)$ resonance
of mass around 1542 MeV,  $\Sigma({3\over 2}^+)$ with mass around
1840 MeV or above, and additional $\Sigma({1\over 2}^+)$ near 1610
MeV.

\begin{acknowledgments}
Helpful discussions with Jia-Jun Wu and Xu Cao are gratefully
acknowledged. This work is supported in part by the National Natural
Science Foundation of China under Grant 10905059, 11035006,
11261130311 (CRC110 by DFG and NSFC), the Chinese Academy of
Sciences under Project No.KJCX2-EW-N01 and the Ministry of Science
and Technology of China (2009CB825200).
\end{acknowledgments}

\newcommand{\etal}{{\em et al.}}

\end{document}